\documentclass[pra,twocolumn]{revtex4} 
\usepackage{graphicx}
\usepackage{bm}
\usepackage{amsmath, amssymb}

\usepackage{color}

\begin{document}

\title{Superconductivity in the two-band Hubbard model}

\author{Akihisa Koga}
\affiliation{Department of Physics, Tokyo Institute of Technology, 
Tokyo 152-8551, Japan}

\author{Philipp Werner}
\affiliation{Department of Physics, University of Fribourg, 
1700 Fribourg, Switzerland}

\date{\today}%

\begin{abstract}
We study the two-band Hubbard model in infinite dimensions by solving the dynamical mean-field equations with 
a strong coupling continuous-time quantum Monte Carlo method 
and show that an 
$s$-wave superconducting state can be stabilized 
in the repulsively interacting case.
We discuss how 
this 
superconducting state competes
with the metallic and paired Mott states.
The effects of the Hund coupling and crystalline electric field are 
also addressed.
\end{abstract}
\maketitle

\section{Introduction}
Strongly correlated electron systems with orbital degrees of freedom 
are attracting much interest in condensed matter physics.
A particularly active topic is the study of 
superconducting (SC) states in transition metal oxides such as 
$\rm Sr_2RuO_4$ \cite{Maeno}, $\rm La(O_{1-x}F_x)FeAs$ \cite{Kamihara}, 
$\rm Na_xCoO_2$ \cite{Takada}, and
fullerene-based solids $\rm A_3C_{60}$ 
(A=alkali metal)~\cite{Hebard,Tanigaki,Ganin}.
These phenomena have been investigated theoretically  
by taking into account
Coulomb interactions, Hund coupling, the multiband structure, 
and the lattice geometry. 
Electron correlations have been treated 
by means of various analytical and numerical methods 
such as the random phase approximation, 
fluctuation-exchange approximation, and dynamical cluster approximation.
While 
the relevant diagrams should be included in the method,
it may be difficult to 
identify which ones are crucial for stabilizing   
the SC state.
A simple question which naturally arises is whether or not a 
SC state can be realized in a model with repulsive interactions, 
without any particular properties of 
the band structure.
Since the onsite Coulomb repulsion usually 
prevents the formation of 
tightly coupled singlet pairs, 
it is expected that an unconventional superconducting state 
with an anisotropic gap should be realized 
instead of the $s$-wave superconducting state.
This 
simple argument 
should be valid in the single band case.

In multiorbital systems, on the other hand, 
the situation may be more complicated. 
Capone {\it et al.} have studied the two-band Hubbard model with 
intraorbital interaction $U$, interorbital interaction $U'$ 
and Hund coupling $J$ \cite{Capone0,Capone}.
They showed that a 
strongly correlated 
SC ground state appears 
under certain conditions when $U=U'$ and $J<0$. 
However, in this analysis, 
the role of the Hund coupling is not quite clear since it is naively 
expected that the interorbital exchange interaction suppresses 
intraorbital pairing correlations.
Therefore, it is instructive to clarify which interactions stabilize 
the superconducting state in the 
two-orbital system. 
Furthermore, it is 
interesting to explore 
whether or not the SC state 
in the strongly correlated region is adiabatically connected 
to the trivial BCS-type SC state stabilized 
by weak attractive interactions.
This may give us a simple explanation 
for the stability of the SC state  
in the repulsively interacting system.

To clarify these issues, we consider the two-orbital Hubbard model.
By combined use of dynamical mean-field theory 
(DMFT)~\cite{Metzner,Muller,Georges,Pruschke} and
the continuous-time quantum Monte Carlo (CTQMC) 
method~\cite{Werner,solver_review},
we show that 
if orbital and magnetic ordering is suppressed, 
an $s$-wave SC state is 
realized 
in the model with degenerate orbitals 
at half filling, and that it is adiabatically connected to 
the simple BCS-type SC state.
The effect of the crystalline electric field is also addressed.

The paper is organized as follows.
In Sec. \ref{2}, we introduce the two-band Hubbard model and 
briefly summarize our theoretical approach.
In Sec. \ref{3}, we demonstrate the existence of a SC state 
in the repulsively interacting region of parameter space.
We study the stability of the $s$-wave SC state 
at low temperatures in Sec. \ref{4}. 
A brief summary is given in the last section.

\section{Model and Method}\label{2}
We consider a correlated electron system with two orbitals, 
which is described by 
the Hubbard 
Hamiltonian~\cite{Momoi,Rozenberg,Held,Han,QMC,Kotliar,Koga1,Koga2,Werner2007}
\begin{eqnarray}
H&=&\sum_{\stackrel{<i,j>}{\alpha,\sigma}}
t_{ij}c_{i\alpha\sigma}^\dag c_{j\alpha\sigma}
-\frac{\Delta_{cef}}{2}\sum_{i\sigma}\Big(n_{i1\sigma}-n_{i2\sigma}\Big)
\nonumber\\
&+&U\sum_{i\alpha}n_{i\alpha\uparrow} n_{i\alpha\downarrow}
+U'\sum_{i\sigma\sigma'}n_{i1\sigma} n_{i2\sigma'}\nonumber\\
&-&\frac{J}{2}\sum_{i}
\Big(n_{i1\uparrow}-n_{i1\downarrow}\Big)
\Big(n_{i2\uparrow}-n_{i2\downarrow}\Big)
\;\;\;
\label{eq:H}\end{eqnarray}
where $\langle i,j\rangle$ denotes 
the summation over the nearest neighbor sites,
$c_{i\alpha\sigma}$ ($c^{\dagger}_{i\alpha\sigma}$) 
is an annihilation (creation) operator
of an electron with spin $\sigma (=\uparrow, \downarrow)$ and
orbital index $\alpha (=1,2)$ at the $i$th site, 
and $n_{i\alpha\sigma}= c^\dag_{i\alpha\sigma}c_{i\alpha\sigma}$.
$t_{ij}$ is the transfer integral,
$\Delta_{cef}$ is the crystal field splitting,
$U$ $(U')$ is the intraband (interband) Coulomb interaction, and
$J$ is the Hund coupling.
Here, we neglect
translational symmetry breaking phases 
such as density waves, antiferromagnetically ordered or  
antiferro-orbital ordered states.
This para-magnetic/orbital solution should be essentially the same as
that for the Hamiltonian 
with next-nearest-neighbor hopping $t'$ equal to 
the nearest-neighbor hopping $t$~\cite{Chitra,Zitzler}.
First, we do not impose any relations between the interactions, 
such as $U=U'+5/2J$. 
This allows us to discuss how the SC state 
which is realized in the single band attractive Hubbard model ($U<0, U'=J=0$) 
is adiabatically connected to
the strongly correlated SC state in model (\ref{eq:H}).
Furthermore, we can discuss how the Hund coupling affects the SC state.
We briefly address the SC state under the condition 
$U=U'+5/2J$ at the end.

To investigate the stability of the $s$-wave SC state, 
we use DMFT~\cite{Metzner,Muller,Georges,Pruschke}.
In DMFT, the original lattice model is mapped to an effective impurity model,
which allows to take into account dynamical correlations. 
The lattice Green's function is obtained via a self-consistency condition 
imposed on the impurity problem. 
The DMFT treatment allows us to discuss the stability of the $s$-wave 
SC state 
more quantitatively than the static BCS mean-field 
theory~\cite{GeorgesZ}.
In fact, the DMFT method has been successfully applied to 
various strongly correlated fermion systems with SC and superfluid
states~\cite{Garg,Dao,Takemori,Bodensiek,KogaSF,Hoshino,Inaba,Okanami}.

When the SC state is treated in the framework of DMFT,
the Green's function should be described in the Nambu formalism.
When the Cooper pairs are formed only in the same orbital,
the impurity Green's function can be written as
\begin{align}
\hat{G}_{\rm{imp},\alpha}(\tau) =
\begin{pmatrix}
G_{\alpha\uparrow}(\tau) & F_\alpha(\tau)\\
F^*_\alpha(\tau) & -G_{\alpha\downarrow}(-\tau)
\end{pmatrix},
\end{align}
where $G_{\alpha\sigma}(\tau)=
-\langle T_\tau f_{\alpha\sigma}(\tau) f_{\alpha\sigma}^\dag(0) \rangle$
denotes the normal Green's function for an electron with spin $\sigma$ and orbital $\alpha$, and 
$F_\alpha(\tau)=-\langle T_\tau f_{\alpha\uparrow}(\tau) 
f_{\alpha\downarrow}(0)\rangle$ and 
$F^*_\alpha(\tau)=-\langle T_\tau f_{\alpha\downarrow}^\dag(\tau) 
f_{\alpha\uparrow}^\dag(0)\rangle$
denote the anomalous Green's functions.
In the calculations, we use a semi-circular density of states,
$\rho(x) = 2 /(\pi D) \sqrt{1-(x/D)^2}$,
where $D$ is the half-bandwidth. 
The self-consistency equation is given by
\begin{align}
\hat{\cal G}_{\rm imp,\alpha}^{-1}&(i\omega_n) = i\omega_n \hat{\sigma}_0 +
\mu\hat{\sigma}_z - \left( \frac{D}{2} \right) ^{2} \hat{\sigma}_z \hat{G}_{\rm imp,\alpha}(i\omega_{n}) \hat{\sigma}_z, 
\end{align}
where $\hat{\sigma}_0$ is the identity matrix and 
$\hat{\sigma}_z$ is the $z$-component of the Pauli matrix, 
$\omega_n =(2n+1)\pi T$ is the Matsubara frequency, and
$T$ is the temperature.
$\hat{\cal G}_{\rm imp}$ and $\hat{G}_{\rm imp}$ are the noninteracting 
and full Green's functions for the effective impurity model.
There are various techniques to solve the effective impurity problem.
A particularly powerful method for exploring finite-temperature properties 
is the hybridization-expansion CTQMC method~\cite{Werner,solver_review}.
It enables us to study the Hubbard model 
 in both the weak- and strong-coupling regimes.
The Green's functions are measured on a grid 
of more than one thousand points.
 
In this paper, we use the half bandwidth $D$ as the unit of energy. 
To examine the nature of the low temperature states,
we calculate the double occupancies $d_{intra}$, $d_p$, $d_a$ and 
the pair potential $\Delta_\alpha$, which are given by 
\begin{eqnarray}
d_{intra}^{(\alpha)}&=&\langle n_{i\alpha\uparrow} n_{i\alpha\downarrow}\rangle,\\
d_p&=&\frac{1}{2}\sum_{\sigma}\langle n_{i1\sigma} n_{i2\sigma}\rangle,\\
d_a&=&\frac{1}{2}\sum_{\sigma}\langle n_{i1\sigma} n_{i2{\bar{\sigma}}}\rangle,\\
\Delta_\alpha&=&\langle c_{i\alpha\uparrow}c_{i\alpha\downarrow}\rangle=
\lim_{\tau\rightarrow 0_+} F_\alpha(\tau).
\end{eqnarray}
We also calculate the quantity $Z_{\alpha\sigma}=[1-{\rm Im} 
\Sigma_{\alpha\sigma}(i\omega_0)/\omega_0]^{-1}$ to estimate the quasi-particle weight 
at finite temperatures. Here, 
$\Sigma_{\alpha\sigma}$ is the normal self-energy 
for spin $\sigma$ and orbital $\alpha$. 

Before we start with discussions, 
let us mention some known facts about the Mott transitions 
in the paramagnetic system. 
When $U'=0$, the system reduces to the single band Hubbard model.
At zero temperature, the transition to the Mott insulating state occurs 
at $U/D\sim 3$~\cite{NRG},
while the pairing transition occurs at $U/D\sim -3$~\cite{CaponePair}.
When $U=U'$, a Fermi liquid behavior is stabilized
up to fairly large interactions due to orbital fluctuations,
and the Mott transition occurs at $U/D\sim 5$~\cite{Koga1,Ono}.
On the other hand, the system is driven to 
the Mott insulating state
away from this condition \cite{Koga1,Koga2}.
In the following, 
we systematically study the appearance of a superconducting phase 
in this model, neglecting translational symmetry breaking phases. 
Then, we clarify the role of the Mott transitions 
in stabilizing the SC state.

\section{Results}\label{3}

\begin{figure}[htb]
\begin{center}
\includegraphics[width=8cm]{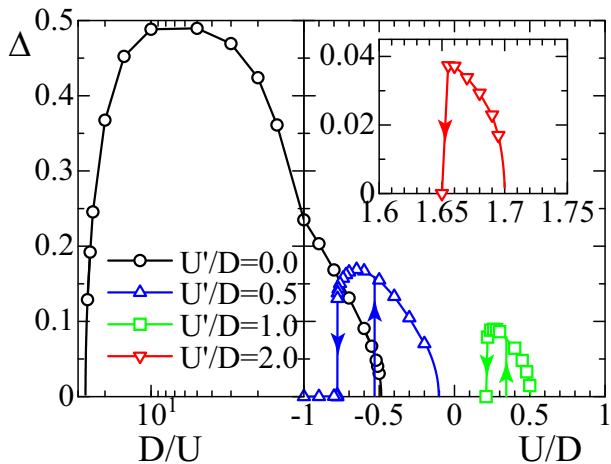}
\caption{(Color online)
Pair potential $\Delta$ as a function of $U/D$ 
when $T/D=0.01$ and $J=\Delta_{cef}=0$.
The arrows indicate the existence of a hysteresis in the pair potential.
The inset shows the results for $U'/D=2.0$.
}
\label{fig:Delta}
\end{center}
\end{figure}

In this section, we discuss the stability of the SC state
in the half-filled two-band Hubbard model with $J=0$ and $\Delta_{cef}=0$.
By combining DMFT with the CTQMC impurity solver,
we obtain the pair potentials at $T/D=0.01$, as shown in Fig.~\ref{fig:Delta}.
When $U'=0$, the two band system is decoupled and becomes equivalent to 
the single band Hubbard model.
In this model, the SC ground state is realized only 
when the intraorbital interaction is attractive. 
At the finite temperature $T/D=0.01$, 
the SC state is realized in the intermediate coupling region
$(U_{c1}<U<U_{c2})$ due to thermal fluctuations, 
where $U_{c1}/D\sim -26.8$ and $U_{c2}/D\sim -0.49$.
In the weak coupling region $U_{c2}<U<0$, the system is metallic. 
Increasing the magnitude of the attractive interaction 
beyond the critical value $U_{c2}$,
weakly coupled Cooper pairs are
formed and the BCS-type SC state is realized.
In the strong coupling region, two fermions are tightly coupled and
composite bosons are condensed, so that a BEC-type SC 
state is realized at low temperatures.
In this case, the energy scale should be $t^2/|U|$,
characteristic of the hopping for paired bosons. 
Therefore, a further increase in the attractive interaction effectively
increases the temperature of the system 
and destabilizes the SC state. 
At $U=U_{c1}$, 
the pair potential vanishes with a second-order phase transition 
to the normal state.
It is known that the BCS-BEC crossover occurs between these regions.
When $U<U_{c1}$, 
the paramagnetic paired Mott (PM) state is realized, 
with an equal probability for  
doubly-occupied and empty states 
at half filling \cite{CaponePair}.

Turning on the interband interaction $U'$
does not simply destabilize the SC state,
but leads to a remarkable behavior, as shown in Fig. \ref{fig:Delta}.
When the onsite interaction $U$ decreases with a fixed $U'/D=0.5$,
the conventional critical behavior appears around $U_{c2}/D\sim -0.10$ 
in the BCS region.
Since the introduction of the interorbital interaction $U'$ increases 
the critical interaction $U_{c2}$, we 
conclude 
that in the BCS region, 
the SC state becomes more stable.
On the other hand, in the BEC region, we find a clear jump singularity
at $U=U_{c1}^{(s)}$ in the curve of the pair potential,
where $U_{c1}^{(s)}/D\sim -0.78$.
This suggests the existence of a first-order phase transition
to the PM state.
In fact, as the intraorbital interaction $U$ is increased 
from the PM state,
the phase transition occurs at $U=U_{c1}^{(m)}$ together with a jump
singularity in the curve of the pair potential,
where $U_{c1}^{(m)}/D\sim -0.53$.
Therefore, the nature of the phase transition in the BEC region is
changed by the introduction of the interorbital interaction,
in contrast to that in the BCS region.
We also find that the first-order phase transition points
are rapidly shifted
to larger $U$, 
as shown in Fig. \ref{fig:Delta}.
This can be explained by the fact that 
the interband Coulomb interactions stabilize 
the PM state, while they have little effect on the SC state. 
When $U'/D=1.0$, we obtain the transition points
$U_{c1}^{(s)}\sim 0.22$, $U_{c1}^{(m)}\sim 0.34$ and $U_{c2}\sim 0.50$.
We note that the $s$-wave SC state is realized 
even in the repulsively interacting region $(U, U'>0)$
although the magnitude of the pair potential is rather small, 
as shown in Fig. \ref{fig:Delta}.
A further increase in the interorbital interaction decreases 
the pair potential until the SC state disappears 
in the large $U'$ region (there is no SC solution 
for $U'/D>3$ at $T/D=0.01$).

\begin{figure}[t]
\begin{center}
\includegraphics[width=7cm]{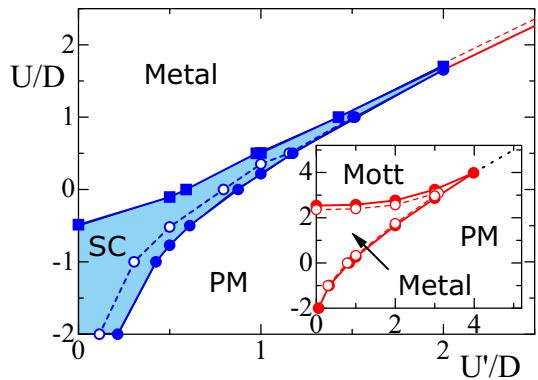}
\caption{(Color online)
Phase diagram of the two-band Hubbard model 
when $T/D=0.01$ and $J=\Delta_{cef}=0$.
Squares represent the second-order phase transition points $(U_{c2})$.
Solid (open) circles represent the transition points 
$U_{c1}^{(s)}\;(U_{c1}^{(m)})$, where the SC (PM) state disappears.
The inset shows the paramagnetic phase diagram of the model.
Solid (open) circles correspond to the transition points where
the metallic (Mott) state disappears. 
The dotted line indicates the crossover line
between the conventional Mott and PM states.
}
\label{fig:PD}
\end{center}
\end{figure}

By performing similar calculations, we obtain the phase diagram at $T/D=0.01$
as shown in Fig. \ref{fig:PD}.
In the negative $U$ region with a small $U'$, 
the trivial SC state
is stabilized by the intraorbital attractive interaction. 
The increase in the interband interaction $U'$ shifts the phase boundaries to larger $U$
and the SC state is stabilized in a narrower parameter region.
The critical end point is located around $(U/D, U'/D)\sim (1.5, 2.0)$.
An important point is that the $s$-wave SC state appears 
even in the repulsively interacting region with $U'>U>0$.
This suggests that the interorbital interactions as well as 
the intraorbital interactions
play a crucial role in stabilizing the SC state in this region.
We note that within the SC region, 
no singularities appear in physical quantities 
such as double occupancies, renormalization factors, or pair potentials.
This 
implies a 
crossover between 
the SC state stabilized by the on-site attraction $U(<0)$
and the SC state for $U'>U>0$. 

A similar behavior has been found in the three-component Hubbard model 
at half filling \cite{Okanami}.
In this case, 
the SC state appears along the first-order phase boundary
between the metallic and PM states in the paramagnetic phase.
However, the paramagnetic properties, and in particular the Mott physics, 
depend on the number of components for each fermion.
For example, in the three component system,
the Mott transition never occurs in the symmetric case $U=U'(>0)$,
while it occurs for the four component system equivalent to 
the degenerate Hubbard model (\ref{eq:H}).
Therefore, it is necessary to clarify how the competition between
the metallic and Mott states in the paramagnetic phase 
affects the stability of the superconducting state.

\begin{figure}[t]
\begin{center}
\includegraphics[width=7cm]{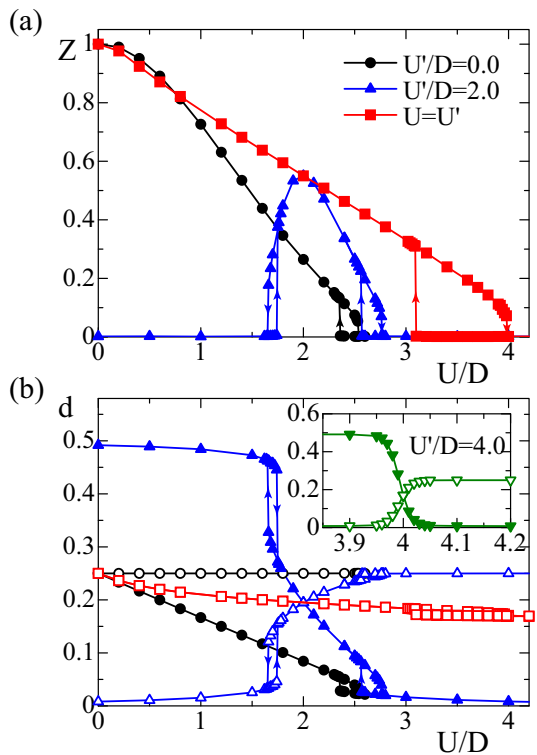}
\caption{(Color online)
Renormalization factor (a) and double occupancy (b) as a
function of the intraorbital interaction $U/D$ 
in the paramagnetic system at $T/D=0.01$ when the interactions 
are fixed as $J=\Delta_{cef}=0.0$ and $U'/D=0.0, 1.0, 2.0$, and $3.0$.
Solid (open) symbols in (b) represent the double occupancy $d_{intra}\; [d_{inter}(=d_p=d_a)]$.
}
\label{fig:Z100}
\end{center}
\end{figure}

For this purpose, we examine the paramagnetic properties 
at the temperature $T/D=0.01$,  
even though the ground-state phase diagram
has already been calculated~\cite{Koga1}.
Figure~\ref{fig:Z100} shows 
the renormalization factors and double occupancies. 
When we increase the intraorbital interaction $U$ with $U'=0$,
the conventional Mott transition occurs around $U/D=2.5$.
In the strong coupling region with $U>U'$, 
$d_{intra}\sim 0$ and $d_p=d_a\sim 0.25$.
This implies that 
each orbital is occupied by one fermion 
and the Mott insulating state is realized.
In the strong coupling region $U/D(>2.8)$ with $U'/D=2.0$, 
a similar behavior appears.
On the other hand, when $U/D<1.6$ and $U'/D=2.0$, 
$d_{intra}\sim 0.5$ and $d_p=d_a\sim 0.0$. 
This indicates that at each site we have a
tightly coupled singlet (empty) state in one orbital,
and an empty (tightly coupled singlet) state in the other. 
Therefore, in this region, the PM state is realized.
In the symmetric case $U=U'$, 
it is known that spin and orbital fluctuations are equally enhanced and 
the metallic state is stable up to a fairly large interaction \cite{Koga1}.
The Mott transition occurs at $U/D \sim 4.0$, where 
$d_{intra}=d_p=d_a\sim 0.167$.
Since the Mott phase transitions are of first order,
a hysteresis appears in these quantities in Fig. \ref{fig:Z100}.
There is a crossover between two Mott states
in the strong coupling region, as shown in the inset of Fig.~\ref{fig:Z100}.

By performing similar calculations in the $(U, U')$ plane,
we obtain the paramagnetic phase diagram of the degenerate Hubbard model,
as shown in the inset of Fig.~\ref{fig:PD}, where
the phase transition from the metallic (Mott) state occurs 
at the phase boundary shown with solid (dashed) lines.
An important point is that the SC state discussed before
is realized along the first-order phase boundary between the metallic and 
PM states 
(Fig.~\ref{fig:PD}). 
This is similar to the finding for the three component system.
Therefore, we can conclude that charge fluctuations for the paired particles,
which are induced by the density-density type interorbital interactions,
play an essential role in stabilizing the SC state.
On the other hand, concerning the critical end point (CEP), 
there is a difference between these two systems.
In the three component systems, 
the CEP for the SC state 
is located near the CEP for the Mott transitions
at low temperatures \cite{Okanami}.
On the other hand, in our model, the conventional Mott state is adiabatically
connected to the PM state through the crossover.
Therefore, at low temperatures, there is no CEP in the paramagnetic state.
This results in the existence of the first order phase transition
between the metallic and PM states
beyond the CEP for the SC state, as shown in Fig.~\ref{fig:PD}.

\begin{figure}[htb]
\begin{center}
\includegraphics[width=6.8cm]{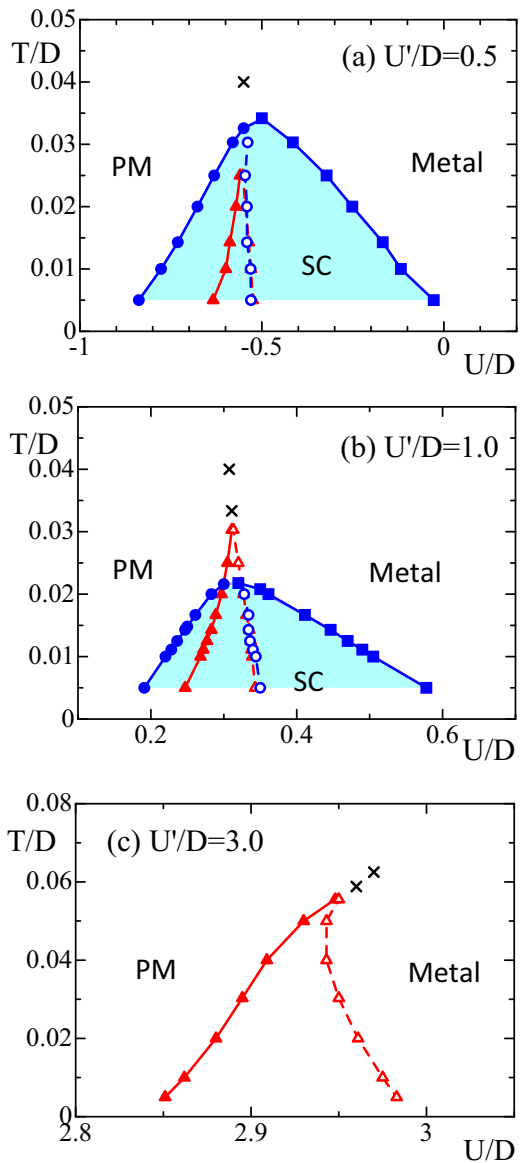}
\caption{(Color online) 
Finite-temperature phase diagrams of the two-band Hubbard model when
$J=\Delta_{cef}=0$ and $U'/D=0.5, 1.0$, and $3.0$.
Squares represent the second-order phase transition points
($U_{c2}$).
Solid (open) circles represent the transition point $U_{c1}^{(s)}$ 
($U_{c1}^{(m)}$) and
solid (open) triangles represent the transition point,
where the metallic (PM) state disappears.
}
\label{fig:PDT}
\end{center}
\end{figure}

This competition leads to interesting finite-temperature properties.
We show the finite-temperature phase diagrams for $U'/D=0.5, 1.0$, and $3.0$ 
in Fig. \ref{fig:PDT}.
In our calculations, it is hard to determine quantitatively
the first-order transition point where the free energies of 
the competing states cross. 
Therefore, we show in the figure 
the coexisting region with competing solutions 
delimited by the open and full symbols.
A crossover between the metallic and Mott states occurs 
at higher temperatures, which is roughly determined 
by the inflection point in the curve of the renormalization factor.
When $U'/D=0.5$, the SC state is realized over a wide $U$-range
at low temperatures due to the strong attractive interaction $U$.
The maximum of the critical temperature is located 
around $(U/D,T/D)\sim (-0.5,0.034)$.
In this case, the Mott phase boundaries for the paramagnetic state 
are within the SC dome, 
as shown in Fig.~\ref{fig:PDT} (a).
Therefore, 
no Mott transition occurs 
and 
the low-temperature properties are similar to 
those for the single band Hubbard model~\cite{Zitzler}.
Increasing the interorbital interaction $U'$, 
the SC region shrinks and the maximum of the critical temperature 
decreases. 
On the other hand, the critical temperature for the Mott phase transition 
shifts to higher temperature. 
This yields a phase diagram with both phase boundaries for 
the Mott and superconducting 
transitions, 
as shown in Fig.~\ref{fig:PDT} (b).
The SC state is realized at low temperatures, 
while the Mott transition occurs for higher temperatures.
We find that 
the Mott 
transition line 
appears 
around the maximum of the superconducting dome.
This is consistent with the observation discussed above, 
that the SC state is realized along the first-order phase boundary 
between the metallic and PM states.
A further increase of the interband interaction destabilizes  
the SC state, so that only
the Mott transition line appears in the phase diagram, 
as shown in Fig. \ref{fig:PDT} (c).
Note that the critical temperature for the Mott transition increases
when the system approaches the SU(4) point ($U=U'$),
which is consistent with the results obtained by 
the self-energy functional theory~\cite{InabaPara}.

Here, we have shown that the $s$-wave SC state 
can be realized in the repulsively interacting model with $U'>U>0$.
It has also been found that the SC state appears 
along the first-order 
phase boundary between the metallic and PM phases.
This observation suggests that the SC state should be stable 
against small perturbations such as a Hund coupling and
a crystalline electric field.
In the following section, we focus on the SC state in 
the repulsively interacting region to discuss its stability.

\section{Stability of the superconducting state}\label{4}

In this section, we consider the effects of 
the Hund coupling and crystalline electric field,
which may affect the stability of 
the SC state in realistic materials.
A positive (negative) Hund coupling favors the parallel (antiparallel) spin 
state between orbitals, while the crystalline electric field favors  
the doubly occupied state in the lower orbital.
Therefore, these terms in the Hamiltonian may destabilize the SC state. 

\begin{figure}[b]
\begin{center}
\includegraphics[width=7cm]{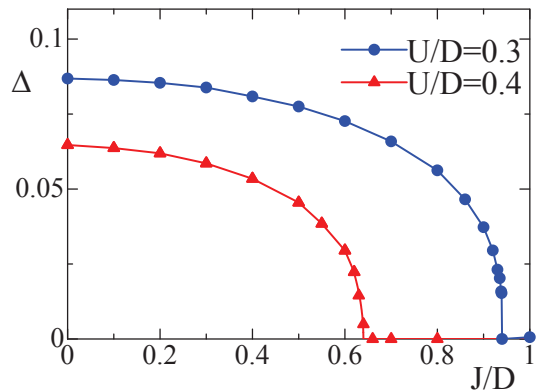}
\caption{(Color online) 
The pair potentials as a function of the Hund coupling $J$
in the two-band Hubbard model at $T/D=0.01$.
Circles (triangles) represent the results for the system 
with $U/D=0.3$ ($0.4$) and $U'/D=1.0$.
}
\label{fig:J}
\end{center}
\end{figure}

First, setting $\Delta_{cef}=0$ and 
fixing the intra- and inter-orbital interactions $U$ and $U'$ 
to certain values,
we discuss how the SC state is affected by 
the Hund coupling.
Note that when the onsite interactions $(U, U',$ and $J)$ 
are independently varied, the pair potential and chemical potential are 
invariant under the transformation $J\rightarrow -J$.
Figure \ref{fig:J} shows the results for the pair potential 
at the temperature $T/D=0.01$.
When $J=0$, the BEC-type (BCS-type) SC state is realized 
with a finite pair potential in the system with $U/D=0.3\;\; (0.4)$ and 
$U'/D=1.0$, as discussed in the previous section.
The introduction of the Hund coupling monotonically decreases 
the pair potential.
Finally, the pair potential vanishes and 
a phase transition occurs to the normal state.
More specifically, when $U/D=0.3\;\; (0.4)$, the phase transition is of first (second)
order, which is consistent with the characteristic feature of 
the BEC (BCS) state.
From the above results, we can say that the Hund coupling simply suppresses 
SC fluctuations.

\begin{figure}[t]
\begin{center}
\includegraphics[width=7cm]{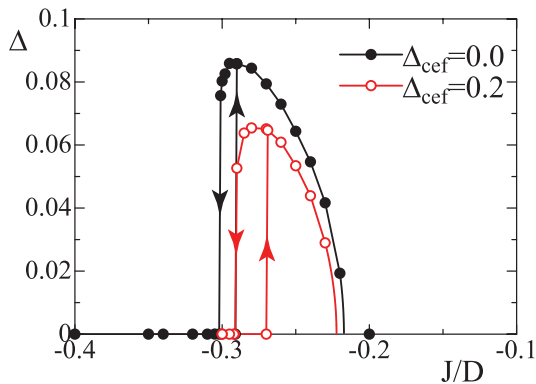}
\caption{(Color online) 
Pair potential as a function of the Hund coupling $J$
in the two-band Hubbard model with the condition $U=U'+5/2J$
when $U/D=0.3$ and $T/D=0.01$.
}
\label{fig:RD}
\end{center}
\end{figure}

On the other hand, if one considers the degenerate Hubbard model 
appropriate for realistic materials,
the onsite interactions $(U, U', J)$ are not independent.
Namely, when the Hund coupling $J$ is varied
with a fixed intraorbital interaction $U$,
the interorbital interaction $U'$ is changed at the same time,
which may induce the SC state.
To clarify this point, we calculate the pair potential in the system under
the condition $U=U'+5/2J$.
When the Hund coupling $J$ is positive,
the interorbital interaction $U'$ is always smaller than
the intraorbital one $U$ and the superconductivity is never realized.
On the other hand, a negative Hund coupling
may drive the system to the SC state since $U'>U$.
The results for the system with $U/D=0.3$ are shown in Fig. \ref{fig:RD}.

First, we focus on the system without crystalline electric field.
When the Hund coupling is small enough $(-0.2<J/D)$,
the system is weakly correlated and the metallic state is realized.
Increasing the magnitude of the Hund coupling,
a second-order phase transition occurs to the SC state
at $J_{c2}/D=-0.22$.
For $J$ below $J_c$, the pair potential grows 
and a SC state is stabilized.
Finally, the first-order phase transition occurs 
at $J_{c1}^{(s)}/D=-0.30$ with
a jump singularity in the pair potential, as shown in Fig.~\ref{fig:RD}.
The double occupancy in each orbital approaches 0.5 and 
the other occupancies $d_p$ and $d_a$ approach zero when $J<J_{c1}^{(s)}$. 
Therefore, the PM state is realized in this region.
The increase of the Hund coupling 
from the PM state induces
another first-order phase transition at $J_{c1}^{(m)}/D=-0.29$.
The obtained results are essentially the same as those discussed 
in the previous section. 
Therefore, we can say that the Hund coupling has little effect 
on the realization of the superconductivity.
The introduction of the crystalline electric field 
increases (decreases) the electron number for the orbital 1 (2),
which suppresses SC fluctuations.
In fact, we find the decrease of the pair potential 
in the intermediate region $(J/D\sim -0.25)$ in Fig. \ref{fig:RD}.

\begin{figure}[t]
\begin{center}
\includegraphics[width=7cm]{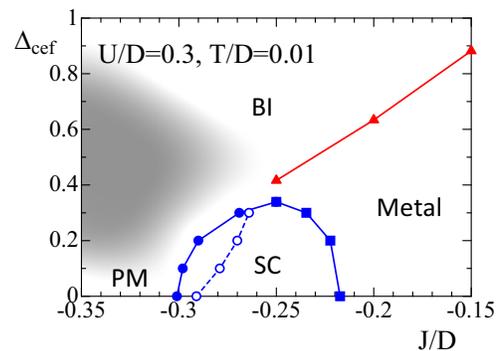}
\caption{(Color online) 
Phase diagram in the space of $J/D$ and $\Delta_{cef}/D$.
Open (solid) circles represent the transition point
$J_{c1}^{(m)}$ ($J_{c1}^{(s)}$) and solid squares represent
the second-order transition point $J_{c2}$.
Solid triangles represent the boundaries between 
the band insulating (BI) and metallic states.
In the shaded area we could not obtain reliable results due to an inefficient Monte Carlo sampling.
}
\label{fig:pd-cef}
\end{center}
\end{figure}

\begin{figure}[b]
\begin{center}
\includegraphics[width=7cm]{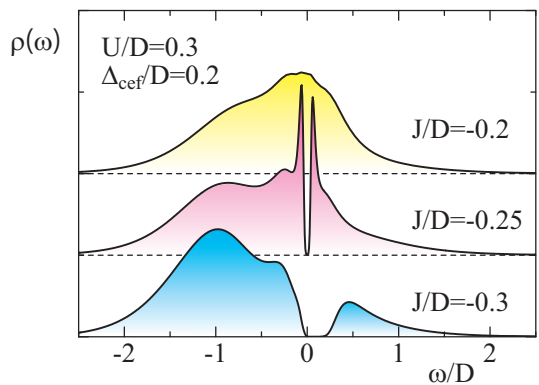}
\caption{(Color online) 
Density of states for electrons in orbital 1 
in the system with $U/D=0.2$ and $\Delta_{cef}/D=0.2$ when 
$J/D=-0.2$, $-0.25$ and $-0.3$.
}
\label{fig:dos}
\end{center}
\end{figure}

By performing similar calculations under the constraint $U=U'+5/2J$, 
we obtain the phase diagram in the space of $J$ and $\Delta_{cef}$, as shown in Fig. \ref{fig:pd-cef}.
One finds that the metallic, SC, and PM states
are stable against a small crystalline electric field.
In the large $\Delta_{cef}$ case,
the lower orbital should be fully occupied leading to a band insulator.
The competition between the PM and band insulating states 
in the paramagnetic region is difficult to pin down  
since the Monte Carlo sampling with standard updates 
becomes inefficient.
Therefore, the boundary between these two states
could not be determined quantitatively.
The $s$-wave SC state appears 
between the metallic, PM, and band insulating states.
Figure~\ref{fig:dos} shows the density of states for the system 
with $U/D=0.3$ and $\Delta_{cef}/D=0.2$ at $T/D=0.01$.
When $J/D=-0.2$, the weakly correlated metallic state 
is realized and no remarkable peak structures appear in the density of states.
When $J/D=-0.25$, a tiny gap appears 
around the Fermi level, which indicates that the BCS-type SC state
is realized.
In the PM state $(J/D=-0.3)$, 
tightly coupled singlet states are formed with $n_1>0.5$ and $n_2<0.5$.
Therefore, a large gap appears in the one-particle spectral function.
This contrast with the conventional Mott insulating state,
which is stabilized only at a commensurate filling.

Before closing this section, we briefly comment on the limitations of our DMFT treatment.
Since this method takes into account local electron correlations, 
we can only discuss the stability of the $s$-wave SC state.
Our calculations have clarified that 
the $s$-wave SC state is realized in the degenerate Hubbard model
when $U'>U$, at least if magnetic and/or orbital ordering is suppressed.  
In transition metal oxides, the intra- and inter-orbital interactions usually satisfy
the relation $U>U'$, meaning that 
it is hard to realize the SC state discussed here.
In this case, 
by taking into account 
the details of the band structure and intersite correlations, 
one has to discuss low-temperature phenomena such as 
unconventional SC with anisotropic gap functions, 
magnetically ordered or orbitally ordered states. 
On the other hand, if one considers low energy states
of systems with a coupling to other degrees of freedom,
the condition $U'>U$ can be effectively realized.
In the fullerene-based solid $\rm A_3C_{60}$, it has been reported that 
this relation should be realized 
due to the strong electron-phonon coupling~\cite{Capone0,Capone,Han2,Nomura}.
In such a situation, DMFT should capture 
the essence of the SC state observed in the material. 

\section{Summary}
We have investigated the low temperature properties 
of the two-band Hubbard model with degenerate orbitals.
By combining dynamical mean-field theory with continuous-time quantum
Monte Carlo simulations, we have 
clarified that a SC state can be realized
in a repulsively interacting two-orbital system due to the competition between
the intra- and interorbital Coulomb interactions.
In particular, this $s$-wave superconducting state appears 
along the first-order phase boundary 
between the metallic and paired Mott states in the para-magnetic/orbital system. 
On the other hand, the Ising-type Hund coupling destabilizes the SC state.  
Although we have not investigated the effect of 
the exchange and pair hopping parts of the Hund coupling term,
we believe that these have little effect 
on the stability of the superconductivity.
It will be interesting to clarify this point in a future investigation.

\section*{Acknowledgments}
The authors thank S. Suga and S. Hoshino for valuable discussions. 
This work was partly supported by the Grant-in-Aid for Scientific Research 
25800193 (A.K.) 
from the Ministry of Education, Culture, Sports, Science 
and Technology (MEXT) of Japan. 
Some of computations in this work has been done using the facilities of
the Supercomputer Center, the Institute for Solid State Physics, 
the University of Tokyo.
The simulations have been performed using some of 
the ALPS libraries~\cite{alps1.3}.

\end{document}